\documentclass{svproc}

\usepackage{color}

\definecolor{hellgrau}{gray}{.70}

\usepackage{tikz}
\usetikzlibrary{
  shapes.symbols,
  positioning
  }

\usepackage[utf8]{inputenc}

\usepackage{todonotes}

\usepackage{placeins}

\usepackage{url}

\usepackage{hyperref}

\title{Leadership Gap in Agile Teams: \newline How Teams and Scrum Masters Mature}

\author{Simone V. Spiegler \inst{1,2} \and Christoph Heinecke \inst{2} \and Stefan Wagner \inst{1}}

	\authorrunning{S. V. Spiegler et al.}
	
	\institute{Institute of Software Technology, University of Stuttgart, Germany\\ \email{\{simone.spiegler,stefan.wagner\}@iste.uni-stuttgart.de}
		\and Robert Bosch Automotive Steering GmbH, Schw\"abisch Gm\"und, Germany\\
		\email{Christoph.Heinecke@bosch.com}}
	
	\titlerunning{How Teams and Scrum Masters Mature}

\begin{document}

\maketitle 

	\begin{abstract}
	\textit{Motivation:}
How immature teams can become agile is a question that puzzles practitioners and researchers alike. Scrum is one method that supports agile working. Empirical research on the Scrum Master role remains scarce and reveals contradicting results. While the Scrum Master role is often centred on one person in rather immature teams, the role is expected to be shared among multiple members in mature teams.\\ \textit{Objective:}
	Therefore, we aim to understand how the Scrum Master role changes while the team matures. \\ \textit{Method:}
We applied Grounded Theory and conducted qualitative interviews with 53 practitioners of 29 software and non-software project teams from Robert Bosch GmbH. \\ \textit{Results:}
	 We discovered that Scrum Masters initially plays nine leadership roles which they transfer to the team while it matures. Roles can be transferred by providing a leadership gap, which allows team members to take on a leadership role, and by providing an internal team environment with communication on equal terms, psychological safety, transparency, shared understanding, shared purpose and self-efficacy. \\
\textit{Conclusion:} The Scrum Master role changes while the team matures.	Trust and freedom to take over a leadership role in teams are essential enablers. Our results support practitioners in implementing agile teams in established companies.
 
	\end{abstract}

\section{Introduction}
\label{intro}

Recently, more and more organisations implement agile teams. Yet, it is not entirely clear how to become agile.
At present, a very popular agile approach is \emph{Scrum} \cite{Schwaber.2017}. Scrum proposes the role of the Scrum Master who takes on a team leadership role \cite{Moe.2010}. The Scrum Master is considered to be a facilitator of the Scrum process and enables a team to work in a self-organised and cross-functional way. Furthermore, the Scrum Master protects the team from external disruptions \cite{Schwaber.2017}.

Yet, empirical research on Scrum teams found that the Scrum Master sometimes acts as a barrier to teams becoming agile in early stages. The reason is that Scrum Masters tend to stick to a command-and-control mode \cite{Moe.2010}. In contrast, teams applying agile methods for three years on average appear not to struggle with the Scrum Master and are even supposed to share the leadership role \cite{Srivastava.2017}.

These diverging results could be explained by changes in the maturity of agile teams. A team learns how to be agile while undergoing different maturity stages \cite{Gren.2017}. Hence, agility of a team is a process that unfolds over time \cite{Werder.2018}. Yet, to the best of our knowledge, there is no empirical analysis of the changing leadership role of the Scrum Master during the agile journey.

To be able to support organisations in the agile transformation, our research objective is to explore the Scrum Master role and how the role changes while the team matures. We believe investigating the changing leadership role of the Scrum Master will provide valuable insights into how teams can become agile.

We collected data in 11 business divisions of the conglomerate Robert Bosch GmbH, primarily operating in the automotive industry. We applied Grounded Theory \cite{Glaser.2017} and conduced qualitative semi-structured interviews with 53 Scrum practitioners from 29 different software and non-software project teams that had applied Scrum over a period of three months up to three years.

We help practitioners in understanding how the Scrum Master can enable a team to become agile in an established company by providing empirical insights on the agile transformation at Robert Bosch GmbH. Applying Role Theory \cite{belbin2012team}\cite{Hoda.2011}, the 53 interviews revealed that Scrum Masters incorporates \emph{nine different roles} which they \emph{transfer to the team while it matures}. We further introduce the concept of a \emph{leadership gap} into research on agile teams which enables team members to take on a leadership role themselves. Hence, we conclude that the Scrum Master role changes while the team matures. 

	\section{Related Work}
	\label{sec:1}

Several authors describe agile teams as being empowered to work in a self-organised and cross-functional manner and that those teams continuously learn and adapt to changing conditions \cite{Conboy.2009}\cite{takeuchi.1986}. Cross-functionality implies to understand each other's roles and domains within one team and to be able to work with each other due to a shared understanding \cite{Hoda.2011}. Self-organisation indicates that teams enjoy a high level of freedom considering how to do their work \cite{cockburn2001agile}. It is no longer the supervisor who assigns tasks to individuals but the team members themselves assign tasks to themselves \cite{cockburn2001agile}. If Scrum teams are led by command-and-control, e.g. members cannot chose their tasks, agility will not materialise \cite{Hoda.2011}.
    
Empirical research on the concept of a Scrum Master shows conflicting results regarding the Scrum Master behaviour. Moe et al. \cite{Moe.2010} find that team members rarely take over responsibility, while Srivastava and Jain \cite{Srivastava.2017} arrive at the conclusion that all team members should be able to take on the Scrum Master role.  
Moe et al. \cite{Moe.2010} observed the implementation of agile methods within one Scrum team over a period of 9 months. They found that the Product Owner and the Scrum Master tend to take over a leadership role most of the time. They describe how a Scrum Master posed a barrier to self-organisation: The person started to control team members which made them stop revealing their impediments and, as a consequence, resulted in weak team leadership and lack of trust. The authors also describe, however, that team leadership improved over time in such a way that more and more team members took over responsibility.
Srivastava and Jain \cite{Srivastava.2017}, who investigated teams that had been working in an agile way for three years on average, outline that the aim of an agile team is to lead themselves. However, they refer to Carson et al. \cite{carson.2007} and acknowledge that taking over responsibility in a Scrum team is a process that unfolds over time due to a shared purpose, social support and voice.

How to evolve from an immature team to a mature one and which role the Scrum Master plays in this journey is yet not clear. We have not found any empirical study on the Scrum Master that specifically examines how the Scrum Master role changes while the team matures.

Cockburn \cite{Cockburn.2002} refers to the Japanese philosophy of Shu-Ha-Ri and describes Scrum as a maturity model for agile adaption. Likewise, Gren et al. \cite{Gren.2017} state that depending on the maturity level of a team, team members practice agile work differently. In the introduction of his doctoral research, Gren \cite{Gren.2017b} suggests that leadership should adapt to different maturity stages of agile teams. He refers to Situational Leadership Theory \cite{Hersey.1979}\cite{Kozlowski.2009} and claims that leaders of agile teams need to demonstrate more monitoring at an early stage but can delegate tasks at a later stage of team development. 
Hoda et al. \cite{hoda.2013} examine both mature and immature teams and discover six different self-organising roles. They claim that roles can be transferred from formal role keepers in rather immature teams to any team member with the right set of skills in more mature teams. 

Since the Scrum Master role has been shown to either display command-and-control behaviour of one formal role keeper in an immature team \cite{Moe.2010} or has been suggested to be played by multiple group members in mature teams \cite{Srivastava.2017}, we believe that the Scrum Master role can be transferred from one individual to distinct team members while the team matures. Yet, this transfer has not been empirically investigated.

	\section{Study Design}
		\label{sec:2}
    
    \subsection{Research Question}		

	Our research objective is to understand the changing leadership role of the Scrum Master in an agile team over time. 

\noindent\textit{RQ1: Which roles does the Scrum Master play in an agile team?}
	
\noindent\textit{RQ2: In which way do team members take on the Scrum Master role over time?}

\noindent\textit{RQ3: How are roles transferred from the Scrum Master to the team members?}
	
\noindent\textit{RQ4: What is the underlying team mechanism required for the role transfer to occur?}

	\subsection{Case and Subject Selection}
    We conducted this study at Robert Bosch GmbH. Two authors have direct access to the field. We identified Scrum practitioners either via our personal network or via intranet and first contacted them by email. We conducted interviews according to availability and willingness to take part. 
    
    We collected data from 11 business divisions which have slightly different subcultures. Most divisions were active in the automotive industry, while others produced domestic appliances and gardening tools. Interviewed teams stated that they apply the method Scrum mostly in modified form, e.g.\ w.r.t.\ the regularity of Scrum meetings. All Scrum Masters were without disciplinary power, responsible for the Scrum process and in charge of team development. Most practitioners of the company call the Scrum Master role \textit{Agile Master}, indicating that this role should adapt to the specific team, rather than sticking to the Scrum approach by the book. Thus, we consider the sample fitting to examine maturity and the changing Scrum Master role.
    
    In total, our data includes 22 Scrum Masters, 8 Product Owners and 23 team members from 14 software development and 15 non-software project teams. The size of teams ranged from 5 to 12 members and often included diverse nationalities. Since the age of teams stretched from three months up to three years, we expected the maturity of teams to vary. To respect participants’ confidentiality, we cite them by SM (Scrum Master), TM (team member) and PO (Product Owner).
	
	\subsection{Data Collection and Analysis Procedure}
	
	To answer our research questions, we conducted semi-structured face-to-face interviews of 45 minutes on average. 
	Interviewees were asked about their personal experiences on agile projects with a focus on the Scrum Master role and what they had learned since they had started to apply the Scrum method. The guiding questions are available online \cite{Spiegler.2018}. 
	Interviews were audio-taped and transcribed. We coded the collected data by applying Glaser's Grounded Theory \cite{Glaser.2017}. We openly coded transcripts sentence-by-sentence and aligned codes that appeared to be alike to one concept. We constantly reflected those concepts critically and aligned them if different concepts appeared to be alike \cite{Glaser.2017}. Through constant comparison \cite{Glaser.2017} of various interviews we identified nine different Scrum Master roles and developed a substantive theory \cite{Glaser.2017} which we labelled the \textit{role transfer process}. 
	
	\subsection{Validity procedure} 
	
	Initially, the sample contained almost only Scrum Masters. Interviews revealed that the Scrum Master role had changed over time and that team members started to take over more responsibility. Drawing on theoretical sampling \cite{Glaser.2017}, we conducted a second round of interviews where we approached the Scrum Master role additionally from the team's perspective and addressed teams as a whole. This showed that team members also had learned to play some of the Scrum Master roles over time. At the beginning of each interview, participants were informed about the purpose of this study and assured of confidentiality, so as to receive open and honest responses. 

	\section{Results}
		\label{sec:3}

Our first two research questions aimed at understanding \textit{which roles the Scrum Master plays in an agile team} (RQ1) and \textit{in which way team members take on the Scrum Master role over time} (RQ2). \\ 
We identified a set of \textbf{nine different roles} that Scrum Masters played. While some teams reported that the leadership roles were rather centred on the Scrum Master, other teams revealed that the Scrum Master role had changed over time. In the latter cases, \textbf{team members started to take over some of the roles} themselves and the Scrum Masters reduced the extent to which they played those roles. In the following Sec.~\ref{nine} we will describe the nine Scrum Master roles. Each role description is divided into three parts:\\
\begin{enumerate}
\item Description of the role in general (RQ1)
\item How the Scrum Master played that role (RQ1)
\item How team members took over that role after some time (RQ2)
\end{enumerate}

\subsection{Nine Scrum Master Roles}
\label{nine}
\subsubsection{Role 1: Method Champion}

Organises meetings and get-togethers, teaches the method, supports formulating tasks and setting goals, visualises information, and discusses how to adapt the method during the retrospective.	

\paragraph{Scrum Master:} A large majority of Scrum Masters mentioned the method to be their main task when working with agile teams. Many emphasised that they continuously helped the team to adapt the method to their specific context.

\paragraph{Team:} In newly established agile teams, members rather waited until the Daily Stand-Up to speak about issues with each other. Over time, teams started to speak with each other right away when an issue occurred. Some teams stated that the Scrum Master had initially organised team events but after some time the team members organised such events themselves. Also, two teams explicitly stated that their team visualised information on a board on their own initiative and that this was the way they learned and exchanged knowledge.

\subsubsection{Role 2: Disciplinizer on Equal Terms}	

Supports the team to keep to the rules, ensures that the team focuses on relevant topics and makes sure that team members attend the meetings. Discipline is accomplished via communication on a par. Interaction on equal terms creates non-hierarchical spaces which are important to speak openly with each other.

\paragraph{Scrum Master:} Initially, some team members were reluctant to follow the Scrum process. When the Scrum Master insisted on discipline, however, such as only talking for a certain amount of time during the daily or to follow up on measurements they had agreed on during the retrospective, the team members started to see the benefit. It is important to note that discipline was described to focus on the Scrum process, not on direct control of team members. If individuals were controlled directly, they reported to loose sense of responsibility. 
    
\paragraph{Team:} Over time, team members learned to focus and prioritise their own work. For example, team members reported to only do one thing at a time and not everything at the same time as they used to do in the past, or they stopped their peers from endless discussions.

\subsubsection{Role 3: Coach}

 Observes team members and uncovers which kind of behaviour is missing in a team to improve teamwork, provides feedback, and helps teams to find out what they wish to change and how to do so.
 
\paragraph{Scrum Master:} Scrum Masters reported to initiate team-building activities, brought developing conflicts to the surface and helped the team to solve them. Coaching was considered important to foster teamwork and  self-organisation.

\paragraph{Team:} Several interviewees described how the Scrum Master built trust among team members, e.g.\ during the retrospective. After some time they established psychological safety \cite{edmondson.1999} and started to open up and to provide feedback to each other. It was no longer merely the Scrum Master who provided feedback to the team.
 
 \begin{quote} \textit{The retrospectives [...] push us to actually stand up for some opinion, to say what is wrong or to open up, and then he [the Scrum Master] unleashed the monster. I have always been very critical about lots of stuff, but now I see that everyone is critical sometimes, now I see that they [the other team members] actually care to say ``look, I am not happy about this'' and speaking openly had never happened before.} \small{(TM)}
 \end{quote}

\subsubsection{Role 4: Change Agent}

Serves as a role model, changes habits, and convinces newly established project teams of the agile way of working.

\paragraph{Scrum Master:} While a large majority helped team members get used to the method step by step, others wanted to help people develop a certain mindset, such as not being afraid of failure or openness towards results. Either way, their overall aim was to convince individuals why the agile way of working made sense. 

\begin{quote} 
\textit{At the beginning, it was a bit tough to convince some team members of the agile approach. But now I think our team does not want to work in a different style anymore. There is a drive in our team that some team members even would like to go further. They have been infected with the agile virus and they want more and more.} \small{(SM)}
\end{quote}

\paragraph{Team:} We did not come across a team members who started to act as a Change Agent pro-actively, such as convincing others of the method. However, several agile teams started to serve as role models for other teams by being agile. The Change Agent role might be important at the beginning of a newly established team. But while the team matures, this role might become obsolete.

\begin{quote} \textit{Back then when I started with agile development, it was rather amusing. Because we felt like animals in a circus. At first, there was astonishment, then amusement, later interest and, finally, they asked whether they [our partner team] couldn't do it the same way. But this was not a process of a few days. It rather took several months.} \small{(PO)} 
\end{quote}

\subsubsection{Role 5: Helicopter}
		
Possesses the ability to see the bigger picture, to know who possess the right skill for a certain task, to include relevant stakeholders and to structure work.
	
 \paragraph{Scrum Master:} They identified cross-boundary links between individuals and tasks from different technical expertise or domains towards a common goal.
	\begin{quote}
		\textit{In our team, I don't feel like I am the boss or anything of that kind. I am just the one in my team who is best at keeping track of things and to give them a general direction.} \small{(SM)}
	\end{quote}

\paragraph{Team:} Due to regular communication and visualisation, team members developed a shared understanding \cite{Moe.2010}\cite{Hoda.2011} while they matured, so that they were aware of who had certain knowledge or skills. Developing a Helicopter perspective helped team members to think in networks, to serve as sparring partners for each other and to be fast in handing over the work to another professional of another expertise.

		\begin{quote}
		    
	\textit{In the beginning, I think you don't know who has more experience in a certain area or expertise in another area. But slowly I get to know everyone and can judge who can support me in which difficulty in the quickest possible way. [...] But in the end I know, okay, I have a problem here and who can support me.} \small{(TM)}
			\end{quote}

\subsubsection{Role 6: Moderator}

Moderates all kind of meetings and builds a bridge between perspectives and domains. This role is considered to be important to develop the necessary cross-functional understanding for agile teamwork.

\paragraph{Scrum Master:} They mediated between individuals from different domains and helped the team to build a shared understanding and to tolerate each other.

\paragraph{Team member:} No interviewee elaborated on a situation in which a member played the Moderator successfully. One team reported that they had tried letting team members lead the retrospective but it had ended up in a planning instead. Two teams felt that the Scrum Master should be the Moderator since they considered it to be difficult to remain neutral during a discussion when being part of the team.

	\subsubsection{Role 7: Networker}

Connects the team with relevant stakeholders, e.g. managers and experts, from within and outside the organisation.

\paragraph{Scrum Master:} The way in which Scrum Masters used their network depended on the current need of the team. Scrum Masters reported that they included formal leaders to gain the support for the agile approach. Another Scrum Master reported how he had invited an expert for a certain method to train the team. Yet another Scrum Master stated that he knew colleagues from facility management whom he could call whenever the team needed organisational support.

\begin{quote}
\textit{You don't have to be better at designing than a design engineer. But you have to somehow show him ways to solve his problems. And if it is only by referring him to another design engineer.} \small{(SM)}
\end{quote}

\paragraph{Team:} The Scrum Master provided contacts and empowered the team members to build their own network over time. This increased their scope of action and enabled them to quickly react to challenges.

\begin{quote} 
\textit{For example, that one has an information for someone, that he normally would not have access to as a planning guy. [\ldots] Actually, I bring in my network from production and the developer his network and the TEF person yet another. During the open discussion at the Daily Stand-Up, I can say that I have a problem. Someone knows someone who can help me with it.} \small{(TM)}
\end{quote}

\subsubsection{Role 8: Knowledge Enabler}
Realises which kind of knowledge the team needs, e.g.\ expert information or methodological skills, and supports team members to acquire that knowledge, e.g.\ sends them to training or conferences, and schedules knowledge exchange meetings. Furthermore, this role promotes iterative learning, e.g.\ learning from mistakes, and fosters learning-by-doing.

\paragraph{Scrum Master:} Some Scrum Masters urged teams to take time for learning. A few of them convinced managers that agile teams must sit close to each other to approach each other easily, learn from each other and build a shared understanding. 
	
\begin{quote} 
\textit{They just do not know the whole approach and how to access it. They know classic learning like you go to a training or you study a book, but in this field, you have so many user groups, meet-ups [\ldots]. And we also try to just propose a nice event. They can meet other people there and discuss with them. For example, we all went to a conference together.} \small{(SM)}
\end{quote}
	
\paragraph{Team:}	While some team members expected the Scrum Master to have the technical expertise to provide feedback, other interviewees had learned to receive feedback from their peers. They shared their progress and served as a sparring partner to each other. They also reported to just walk over to colleagues and ask for information, sit together when they had questions or collaborate on tasks. 
	
\begin{quote} 
\textit{Today it is all very easy going. I just go over to my colleague's desk, sit down for, like, 45 minutes and work with him on a topic. Nobody says anything against that. It is very informal, but it also happens that I personally have to answer some questions. 
} \small{(TM)}
\end{quote}

\subsubsection{Role 9: Protector}
Shelters teams from inappropriate requests from the Product Owner, managers, disciplinary leaders and other departments.
	
\paragraph{Scrum Master:} Scrum Masters reported protecting the team from re-prioritisa\-tion or too high workload by the Product Owner. Furthermore, they sheltered the team from management intervening in daily business or overruling decisions the team had come up with.
		
	\begin{quote} 
	\textit{But then, I also pushed some things through in certain teams, [...] in which managers had taken decisions again. I had to go to the management and tell them ``that is not OK, you make a mistake''. Then they had to compromise and later they were really glad that they had reacted that way. Because the team gave the right hints after all. That is a situation in which one has to fight a battle on behalf of the team.} \small{(SM)}
	\end{quote} 
		
\paragraph{Team:} One team implemented a role called ``Batman'' that was responsible for the protection from external requests that were unrelated to the respective sprint goal. The role keeper changed depending on day and time. Two teams reported that they struggled because the Scrum Master currently had no time to stick with the team regularly. One team stated that it had happened twice that management removed members temporarily during the absence of the Product Owner. The Product Owner wished that there was a Scrum Master on a regular basis to defend the team. Likewise, a Product Owner of another team said that he struggled with not intervening in operational work and tended to tell people what to do. He wished that there was a Scrum Master regularly to stop him from disturbing. Thus, we suggest that it might be difficult for teams to protect each other from management in an established company.

Investigating which role the Scrum Master plays (RQ1) and in which way it changes over time (RQ2), we found that \textbf{the Scrum Master played nine different roles which he transferred to the team while it matured.} In addition, we found  that some roles were more suitable for a transfer to the team members than others. In the following, we will elaborate on how roles were transferred from the Scrum Master to the team members.

\subsection{The Role Transfer Process}

\begin{figure}
\begin{tikzpicture}[
  pfeil/.style={draw,
  signal,signal pointer angle=120,signal from=west,
  text width=2.9cm,minimum height=0.7cm,inner xsep=1em},
  scale=.79,transform shape,
  node distance=.01cm]
\node[pfeil](pfeil1){\textbf{Scrum Master}};
\node[pfeil,right=of pfeil1](pfeil2){demonstrate role};
\node[pfeil,right=of pfeil2](pfeil3){leadership gap};
\node[pfeil,right=of pfeil3](pfeil4){support if needed};
\end{tikzpicture}

\begin{tikzpicture}[
  pfeil/.style={draw,
  signal,signal pointer angle=120,signal from=west,
  text width=2.9cm,minimum height=0.7cm,inner xsep=1em},
  scale=.79,transform shape,
  node distance=.01cm]
\node[pfeil](pfeil1){\textbf{Team Member}};
\node[pfeil,right=of pfeil1](pfeil2){observe role};
\node[pfeil,right=of pfeil2](pfeil3){claim and grant role};
\node[pfeil,right=of pfeil3](pfeil4){play role};
\end{tikzpicture}

	\caption{The three steps of the role transfer process.}
		\label{fig:1}
	
	\end{figure}
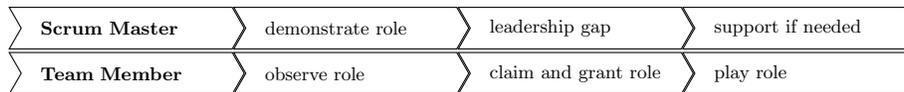

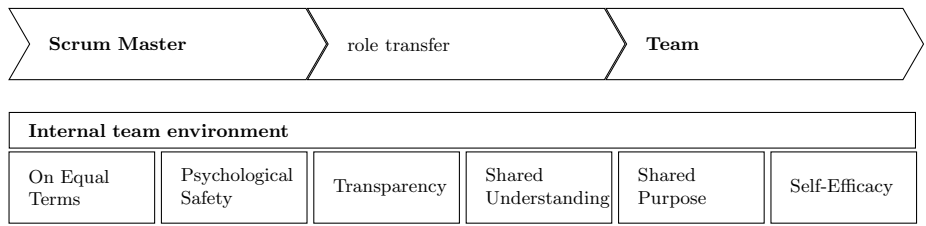
\begin{figure}

\begin{tikzpicture}[
  pfeil/.style={draw,
  signal,signal pointer angle=120,signal from=west,
  text width=4.0cm,minimum height=1.2cm,inner xsep=1em},
  scale=.79,transform shape,
  node distance=.01cm]
\node[pfeil](pfeil1){\textbf{Scrum Master}};
\node[pfeil,right=of pfeil1](pfeil2){role transfer};
\node[pfeil,right=of pfeil2](pfeil3){\textbf{Team}};
\end{tikzpicture}
\\

\begin{tikzpicture}[
  pfeil/.style={draw,
  signal,signal pointer angle=180,signal from=west,
  text width=14.6cm,minimum height=0.6cm,inner xsep=1em},
  scale=.79,transform shape,
  node distance=.1cm]
\node[pfeil](pfeil1){{\textbf{Internal team environment}}};
\end{tikzpicture}

\begin{tikzpicture}[
  pfeil/.style={draw,
  signal,signal pointer angle=180,signal from=west,
  text width=1.8cm,minimum height=1.2cm,inner xsep=1em},
  scale=.79,transform shape,
  node distance=.1cm]
\node[pfeil](pfeil1){{On Equal Terms}};
\node[pfeil,right=of pfeil1](pfeil2){Psychological\\Safety};
\node[pfeil,right=of pfeil2](pfeil3){Transparency};
\node[pfeil,right=of pfeil3](pfeil4){Shared\\ Understanding};
\node[pfeil,right=of pfeil4](pfeil5){Shared\\ Purpose};
\node[pfeil,right=of pfeil5](pfeil6){Self-Efficacy};
\end{tikzpicture}

\caption{Integrative model of the role transfer process.}
 \label{fig2: 2}
\end{figure}

	    Our third research question was: \textit{How are roles transferred from the Scrum Master to the team members?} We found that roles were transferred via \textbf{three steps} we labelled the \textbf{role transfer process} (shown in figures \ref{fig:1} and \ref{fig2: 2}).  Before we elaborate on the role transfer process in depth, we exemplify the concept by referring to the following story:
	    
	   When asked how he had learned to take over responsibility, one team member said that he had faced a major challenge in an area in which he had no previous experience. He had encountered a lack of leadership since no expert was there to support him.  He said that, initially, he was afraid of taking over the responsibility necessary to tackle the challenge. Yet, he was left alone and had to solve the challenge himself. He had felt a sense of personal responsibility. Consequently, he had decided to take over a leading role. He felt a high level of \textbf{self-efficacy}\label{self-efficacy} after he had solved the challenge successfully and became a very proactive team member afterwards. He stated that in his current project, he missed such a lack of leadership and that he had observed that team members were reluctant to self-assign tasks. We name this lack of leadership which provides the opportunity for a team to take on a leadership role the \textbf{leadership gap}. 
	    
	\FloatBarrier

\textbf{We found that the role transfer process consists of three steps:}

\textbf{The first step} describes how the Scrum Master serves as a role model by performing all nine roles. The Scrum Master demonstrates how to perform the activities of the roles, while team members observe and communicate regularly on the meaning of the roles, e.g.\ with the help of visualisation. They build a \textbf{shared understanding}\label{shared understanding} concerning the Scrum Master role which leads to role clarity.

\begin{quote}

\textit{If you want to do Scrum, you have to make sure that people understand the different roles.} \small{(SM)}
\end{quote}

\textbf{The second step} describes that Scrum Masters stop playing certain roles themselves after some time and simultaneously prevent management and Product Owners from taking on the respective role. While some teams stated that management allowed them to actually decide, others experienced that management, Scrum Masters or Product Owners were reluctant to hand over power and, consequently, team members did not take over leadership roles. Contrarily, team members who face a leadership gap in which no one plays the specific role get the opportunity to take on the roles themselves. If a team member claims the leadership role for him- or herself, other team members allow the respective team member to take over that role and accept the new role keeper. \\
 
\begin{quote}
    
\textit{As a Scrum Master I can provide strong support at the beginning to get started. But then I have to retreat gradually so that the team gets into the mode of self-organisation. Because if you do not create some free space or a vacuum, nobody will jump in.} \small{(SM)} 
\end{quote}

\begin{quote}
\textit{I try to help colleagues to find their way into the roles. It is always tricky to keep the balance between what the team should do by themselves and what should be done by the PO or SM. That is one thing that one has to reflect upon and to level out one. [...] The most exciting thing is to bear the silence until someone says something and to wait until someone else gets active. [...] Also we have to give them some free space to experiment and try out themselves.} \small{(SM)}
\end{quote}

Scrum Masters stated that they either provided a leadership gap on purpose by not playing certain roles but waiting that team members would take on the opportunity and play the role, or they were not playing the role because they were absent which gave the team the chance to perform the role.

\begin{quote}
\textit{I did not have sufficient capacity to do everything myself. Therefore, some team members took over tasks, e.g.\ one guy arranged a timer, another one took care of the whiteboard. They were quite proactive as a team. [..] They did not tell me: ``You are in the Scrum Master role, you have to make things better for us.''} \small{(SM)}

\end{quote}

\textbf{The third step} describes that team members play most of the roles while the Scrum Master only continues to perform a role when still needed. The Moderator and Protector roles were found to be difficult to be transferred to the team, e.g.\ because the role keeper should remain unbiased. This indicates that the Scrum Master role does not become obsolete but is played to a lesser extent by a formally appointed person over time.
Therefore, we suggest that some roles should always be played by the Scrum Master, which is a similar result as the findings by Hoda et al.~\cite{hoda.2013} who discovered that in the absence of specific formal role keepers some aspects of agile working lost the team's attention, such as the retrospective.

\begin{quote}
    
\textit{It takes a lot of energy but is quite nice to experience when the team gradually walks by itself. At the same time, the time effort by the Scrum Master can be reduced.} \small{(AM)}
\end{quote}

 \subsection{Internal Team Environment}
 
Our fourth research question aimed at understanding \textit{the underlying team mechanism required for the role transfer to occur.} We found \textbf{six patterns} shaping an \textbf{internal team environment} that supported team members to take on leadership roles. As already exemplified at the beginning of Sec.~\ref{self-efficacy}, we found that \textbf{self-efficacy} increased when performing leadership roles and stimulated team members even further to continue playing those roles in the future. We also found that it is important to develop a \textbf{shared understanding} \cite{Moe.2010} concerning the roles which was often reached by \textbf{transparency}, e.g.\ during the retrospective.

Additionally, the Discipliniser on Equal Terms role described that teams communicated with each other \textbf{on equal terms}. This means that team members have to be free from hierarchical thinking to claim and grant a leadership role. The overall atmosphere within the team should reflect that it is generally accepted to take on a leadership role even without being a formal leader of the group. As a result, claiming and granting of leadership roles will occur.

Furthermore, team members have to establish \textbf{psychological safety} \cite{edmondson.1999}\cite{Moe.2010} to feel safe taking over the risk of playing a leadership role without previous experience in it. The Scrum Master was found to provide safety by the Scrum process. The retrospective helped teams to build trust among each other and encouraged team members to talk openly about personal matters. This fostered a feeling of safety within the group. 

\begin{quote}
\textit{I think, first, one benefit is that we learned [\ldots] to lose the fear of talking. So the methods forced us to bring out opinions, to give the opinions on something that was bothering.} (TM) 
\end{quote}

Moreover, \textbf{shared purpose} is important \cite{carson.2007}\cite{cockburn2001agile} for team members to feel responsible. One team struggled in iterative learning and they complained that they had no vision. They felt like the knowledge they were required to learn was useless, and they did not understand why they should learn continuously. As a consequence, the team members seldom took over the Knowledge Enabler role. Based on our empirical evidence we assume that teams need a shared purpose to be willing to take on a leadership role.

Answering our third (RQ3) and fourth research questions (RQ4), we found that \textbf{an internal team environment} of self-efficacy, shared understanding, transparency, communication on equal terms, psychological safety and shared purpose
\textbf{enabled roles from the Scrum Master to be transferred to team members during the role transfer process.}

\section{Discussion and Relation to Existing Evidence}	
  	\label{sec:4}
Our research objective was to explore how the Scrum Master role changes while the team matures. We discovered that the Scrum Master comprises nine leadership roles. While the team matures, more and more roles are transferred from the Scrum Master to the team. At the heart of the role transfer process lies the leadership gap: a lack of leadership which provides the opportunity for team members to step up and take on leadership roles which were previously filled by the Scrum Master.

Several authors found that interference from Scrum Masters, Product Owners or management decreased self-organisation of teams \cite{Moe.2010}\cite{Hoda.2011}, while communication among team members improved when the Scrum Master was absent \cite{Moe.2010}. We believe that our finding of providing a leadership gap that allows teams to take over leadership roles fits well with those earlier observations. 
	 
Furthermore, the Scrum process, e.g.\ retrospectives, stimulates a certain internal team environment \cite{Hoda.2011} which empowers team members to take on leadership roles. While recent research urged practitioners to focus on human interaction instead of the method \cite{Werder.2018}, we claim that it is not an either-or approach but that human interaction and the method go hand in hand.	Based on our results, we argue that the Scrum method combined with a certain behaviour, such as communication on equal terms, fosters a supportive team environment, such as mutual understanding and trust \cite{Moe.2010}. This empowers teams to take on leadership roles while they mature.

	\section{Practical Implications} 
	\label{sec:5}	

Many practitioners on the management level have set the agile transformation of their organisations as one of their top priorities. This is often associated with the common misconception that when implementing agile projects their teams are instantly ``doing twice the work in half the time'' as the famous title of the book on the Scrum method promises \cite{sutherland2014scrum}. Few have understood and accepted the time required for the team development process.

When agile teams are implemented in established companies, individuals have to learn a new way of leadership in teams, which will lead to slower delivery of work products at the beginning. Management should grant sufficient time to teams to allow them to regularly reflect upon the leadership roles during the retrospective, learn their meaning and content, build a mutual understanding and figure out how and to what extent to take on leadership roles.  Teams need time to try the roles and learn them, possibly by failure. Just like any newbie in a formal leadership position needs time and is given time to learn the role, agile teams need time to learn the leadership roles of Scrum.

Furthermore, even though management expects employees to change and take on more responsibility, some managers are reluctant to grant leadership roles to the teams. External pressure, top-down changed targets and shifted priorities as well as frequent changes of the team setup destroy the sheltered space within which agile teams can grow. In established companies, it is easy to re-staff project teams because authorities have the legitimate power to do so, but it is a supreme discipline to protect the team and create hierarchcy-free space for team development by the Scrum Master demonstrating lateral leadership.

Therefore, management must provide a Scrum Master to protect the team and shelter it while it matures. The Scrum Master has to be granted sufficient managerial power to protect the team and to preserve the leadership gap as a major enabler for the team’s transformation.
Simultaneously, the Scrum Master must be patient and wait until team members take on responsibility when they face a lack of leadership. Likewise, team members  have to learn how to practice new ways of interacting with their team and managers, and to develop the courage to bridge the leadership gap when provided, even though it might feel inconvenient at the beginning.

\section{Limitations and Future Work}
	\label{sec:6}

To assure the quality of our research, we critically discuss construct validity, external validity and reliability: \\
To increase \textbf{construct validity}, we used multiple sources of evidence by capturing the Scrum Master role from three different angles involving Scrum Masters, Product Owners and team members. The researchers discussed the extracted results and built concepts and theories. Additionally, emerging results were frequently reflected critically with various agile practitioners from the company. Furthermore, the main author observed multiple agile teams at the company site over a period of 1.5 years. The final results were supported by the observations and fruitful discussions with practitioners. 

All participants work at the same conglomerate, mostly in the automotive industry. To increase \textbf{external validity}, we tried to ask an equal number of project teams at each division. Despite their slightly similar overall working culture, the 11 business divisions embrace different subcultures. Still, we do not claim our results to be universally applicable and they might be limited to the specific context. Further studies should compare our findings on the changing Scrum Master role with the results emerging from other conglomerates.

\textbf{Reliability}: Since we used an open-ended semi-structured questionnaire that guided us through the interviews, the different interviews followed a similar structure. Yet, we asked participants about past events and what they had learned over time. Memories of individuals tend to change in retrospective. Therefore, these interviews are difficult to replicate. A cross-sectional follow-up survey on a nominal scale containing the nine roles with the respective activities we have identified would increase the reliability of this study.

 We have not yet captured the perspective of the management who may experience providing the leadership gap differently. Taking this aspect into account, we would like to extend this study in the near future and dig deeper into strategies on how the Scrum Master protects the leadership gap from management and Product Owner by referring to the boundary-spanning role of team leadership literature \cite{druskat2003managing}.

\bibliography{references.bib}
  
		%
\end{document}